# Online Product Feature Recommendations with Interpretable Machine Learning

**Mingming Guo**
The Home Depot
Atlanta, GA, USA
mingming_guo@homedepot.com

**Nian Yan**
The Home Depot
Atlanta, GA, USA
nian_yan@homedepot.com

**Xiquan Cui**
The Home Depot
Atlanta, GA, USA
xiquan_cui@homedepot.com

**Simon Hughes**
The Home Depot
Chicago, IL, USA
simon_hughes@homedepot.com

**Khalifeh Al Jadda**
The Home Depot
Atlanta, GA, USA
khalifeh_al_jadda@homedepot.com

## ABSTRACT

Product feature recommendations are critical for online customers to purchase the right products based on the right features. For a customer, selecting the product that has the best trade-off between price and functionality is a time-consuming step in an online shopping experience, and customers can be overwhelmed by the available choices. However, determining the set of product features that most differentiate a particular product is still an open question in online recommender systems. In this paper, we focus on using interpretable machine learning methods to tackle this problem. First, we identify this unique product feature recommendation problem from a business perspective on a major US e-commerce site. Second, we formulate the problem into a price-driven supervised learning problem to discover the product features that could best explain the price of a product in a given product category. We build machine learning models with a model-agnostic method *Shapley Values* to understand the importance of each feature, rank and recommend the most essential features. Third, we leverage human experts to evaluate its relevancy. The results show that our method is superior to a strong baseline method based on customer behavior and significantly boosts the coverage by 45%. Finally, our proposed method shows comparable conversion rate against the baseline in online A/B tests.

## CCS CONCEPTS

• **Information systems** → Recommendation systems; Information retrieval; • **Computing methodologies** → Machine learning.

## KEYWORDS

Recommendation systems, feature importance, ranking features, tree-based models, linear models, shapley values, product price, left navigation algorithm, customer behaviors.

**ACM Reference Format:**
Mingming Guo, Nian Yan, Xiquan Cui, Simon Hughes, and Khalifeh Al Jadda. 2020. Online Product Feature Recommendations with Interpretable Machine Learning. In *Proceedings of ACM SIGIR Workshop on eCommerce (SIGIR eCom'20).* ACM, New York, NY, USA, 7 pages.

## 1 INTRODUCTION

E-commerce companies rely on recommendation systems to assist customer shopping in the online setting and create pleasant shopping experiences using machine learning. Alternative product recommendations provide similar product options and help customers find the product that they like from a massive online catalog. When customers compare alternative products, product features are essential for them to find the differences among those alternative products and choose the product with features that fit their needs. Product feature recommendation is an emerging topic to achieve this purpose. By showing the right feature set, customers can quickly figure out the desired feature configurations. One example of product feature recommendations is shown in Figure 1. When a customer wants to buy 'Patio Dining Sets', 'Seating Capacity' is one of recommended key product features for the customer to make the right purchase decision between the two alternative products.

**Figure 1: Product Feature Recommendations**

However, from a major US e-commerce site, we found that the features shown for alternative products are not very helpful when customers are making purchase decisions. Thus, we need to understand the key differences between product features and what are the major factors that drive the difference between alternative products. The task of product feature recommendations is to recommend differential features considering those factors. From the literature, a lot of research has been focused on product recommendations such as alternative product recommendation [9, 14], complementary product recommendation [3], and/or next item recommendation [13]. Little work has focused on product feature recommendations. Feature recommendations help customers to interpret product recommendations by showing the key differences. Making product recommendations interpretable to customers is important to gain trust from customers, boost shopping confidence, and increase the purchase probability [16].

In this paper, we propose an interpretable machine learning approach to generate feature recommendations that drive the major differences between alternative products. To be specific, we propose to use the product price as the training label and build machine learning models to learn the feature importance and impact (positive or negative) for a given product category. We then study the *Shapley Values* [5] with visualization for interpreting the feature contributions to the product price from the models and obtaining the final feature recommendations. Our main contributions are as follows:

- Identify the unique product feature recommendation problem to help customers make better purchase decisions between alternative products by showing top ranking features.
- Formulate the problem into a price-driven supervised learning problem to derive the most important product features per category.
- Build interpretable machine learning models to understand the feature importance and whether features have a positive or negative impact on the product price.
- Utilize model-agnostic method *Shapley Values* to interpret the contribution of the features to the models' predictions based on rich data visualizations and finally rank the features per category.
- We leverage human experts to label the top features for sampled categories and evaluate the proposed method against a strong baseline named *Left Navigation Algorithm* that uses customer behaviors.
- Our proposed approach scores higher in the offline evaluation on metrics such as NDCG, precision, recall and coverage; it's also comparable with the *Left Navigation Algorithm* in online A/B tests on conversion rate.

## 2 RELATED WORK

A lot of research has been conducted into online recommendation systems from both an academic and industrial setting [15]. Among this work, product recommendation systems attract more attention due to their wide usage in online services such as social media and e-commerce sites. However, product feature recommendation is an emerging topic in recommendation system research due to its specific settings in e-commerce.

### 2.1 Product and Feature Recommendations

For product recommendations, content-based recommendation is an approach for recommending products/items based on the content similarity, and collaborative filtering is another popular approach which leverages user behaviors instead of content to generate recommendations for specific users or products [6]. Deep learning is also being heavily applied for recommendation systems [15]. Despite the active research into recommendation systems in general, little work has been carried out into building product feature recommendation systems. This problem is essential to customers because it can help differentiate specific products from a set of similar products based on the recommended features. In the literature, recommendation of product features or attributes has been used for designing marketing campaign [12].

### 2.2 Interpretable Machine Learning

*Interpretable Machine Learning* tries to explain why and how a machine learning model works to facilitate human understanding of the final model [10]. Thera are two major techniques in it: interpretable models and model-agnostic methods.

*2.2.1 Interpretable Models.* For interpretable models, *Linear Regression* [1] is one of them that can clearly explain the linearity relationship between the features and the target variable. However, it fails in the situation where there is a nonlinear relation between features and targets or the features are correlated to each other. *Decision Tree* [2] is another model with interpretability. It splits the data with certain cutoff values from the features to build the tree in an iterative way. Based on *Decision Tree*, there are many advanced tree-based models including *LightGBM* [4] and *CatBoost* [11].

*2.2.2 Model-Agnostic Methods.* Model-agnostic methods are methods that apply on top of machine learning models that helps humans understand. *LIME* [8] is a model-agnostic method to explain the predictions in machine learning. However, this method focuses on machine learning classifier but not regression tasks. SHAP [7] which stands for *SHapley Additive exPlanations* is a universal game theory method to interpret the feature importance for machine learning models in prediction tasks. It can obtain the *Shapley Values* for the features in the prediction process from the training data. There are also other model-agnostic methods including Feature Interaction, Permutation Feature Importance, etc [10]. Based on the model-agnostic capacity and rich visualization, we select SHAP as the tool to interpret machine learning models and understand the feature importance for our problem.

## 3 MACHINE LEARNING APPROACH

In this section, we describe the characteristics of the dataset and how we process the raw data and extract features. We build regression models using *Linear Regression*, *LightGBM* and *CatBoost* to learn and understand the feature importance using product prices as labels. We also leverage *Linear Regression* to learn the feature directions. We further study *Shapley Values* to interpret the feature contributions from each model and compute the final feature ranking list by averaging the *Shapley Values* for each feature from all three regression models. We learn the feature importance to understand what features contribute more to the product's price, and use the feature direction to understand if a feature positively or negatively contributes to the product price.

## 3.1 Datasets

In our catalog, there are millions of online products which are organized by categories. A category is defined as a classification for certain type of products. For example, 'hammer' is a specific category within which there are all hammers with different brands and prices. We focus on leaf node level category from the classification tree since the products in each leaf node category are similar products and sharing similar features. To develop the models, we do random sampling and select seven categories from our catalog. We need to accurately extract the most important features and their directions for each category. An example of the raw product attribute data for an online product in the 'Dehumidifiers' category is shown in Figure 2. We use feature and attribute interchangeable in this paper. For each category, the number of products is ranging from hundreds to tens of thousands. The number of features for each product in each category is ranging approximately from 5 to 30.

| displayname | attributevalue |
|---|---|
| Package Size | Individual |
| Commercial/Residential Use | Residential |
| Product Weight (lb.) | 51.8 lb |
| Product Width (in.) | 16.8 in |
| Bucket capacity (pints) | 13.3 |

**Figure 2: An example of raw product feature data**

## 3.2 Data Pre-Processing

We process the data by converting the raw string data into different types of features. There are 3 types of features we need to extract in the raw data including numerical features, categorical features and textual features. The detailed processing methods for each type of feature are as follows:

- **Numerical Features**: from Figure 2, a feature like 'Product Weight(lb.)' has a unit 'lb' in the raw data, so we need explore the raw data to discover such patterns. We apply the patterns/rules with regular expressions to remove information such as unit and extract the numerical part of the features.
- **Categorical Features**: we also extract features like 'Commercial/Residential Use' with short string value 'Residential' as a categorical feature shown in Figure 2. We also find that there are raw features with long string information like some functional description. Based on the statistics, we choose maximum 10 different strings in a feature column as a categorical feature.
- **Textual Features**: from the process to extract categorical features, we find that there are some features with lots of different long string/textual values (e.g., describe the product usage information) which are not informative so we exclude those features. Those features are not very useful for customers to differentiate the products compared with numerical features like 'Bucket capacity (pints)' or categorical ones like 'Product Width (in.)'.

## 3.3 ML Algorithms

**Linear Regression**: *Linear Regression* is the first machine learning algorithm we use to understand the feature importance and feature direction due to its interpretability. The basic idea is that each feature is assigned a weight and the weight is updated in the training process to minimize the prediction error. We also focus on *Linear Regression* to learn the positive or negative sign of each feature, which means if the increase or decrease of a feature value will make the product price higher. *Linear Regression* can learn the impact direction of a feature by increasing or decreasing the weight for that feature while observing the changing direction of the target variable like price. When changing the weight for a feature, the weights of other features should be kept constant without changing. These weights are the feature importance. For each categorical feature, we sum the weights from the dummy features generated by one-hot encoding from that feature to get a single score. The feature sign is important for explaining the price changing direction when comparing two products for a specific feature. This serves another business case in the Section 4.4.

**LightGBM**: *LightGBM* is a gradient boosting tree-based algorithm we choose because it performs well in many prediction tasks with faster training speed and better accuracy in general. This is the first tree model we use to learn and understand the feature importance. The way for *LightGBM* to learn the feature importance is by computing the average gain of the features when they are used to partition the data during the model training. We also need to convert categorical features into numerical format using one-hot encoding. Similarly, for each categorical feature, we sum the scores from the dummy features generated by one-hot encoding from that feature to get a single score. We use the model result from *LightGBM* as an input to the final feature ranking process.

**CatBoost**: *CatBoost* is another sophisticated tree-based algorithm that can better handle categorical features directly as a whole without the need to split them. This is the major reason we choose it. *CatBoost* supports textual features as well. The way for *CatBoost* to learn feature importance is also based the average gain of the features in the tree splitting process. Specifically, *CatBoost* learns the feature importance for categorical features by transforming them into numerical before each split is selected in the tree and using various statistics on the combinations of categorical and numerical features as well as combinations of categorical features in the process. We also use the model result from *CatBoost* as an input to the final feature ranking process.

## 3.4 Shapley Values

We mainly study the *Shapley Values* to better interpret the feature contributions for the machine learning models we have built and obtain the final ranking features per category. The major advantage of *Shapley Values* is that it is an explanation method with a solid theory [10]. *Shapley Values* is a method developed in colitional game theory. In a prediction task, considering each feature value of an instance as a player and the payout as the prediction value, *Shapley Values* is about how the payout is distributed to the features or players in that prediction. Specifically, for a single feature/player from all training instances, its marginal contributions with all possible coalitions with other features/players can be calculated based on the differences between the predicted value for that instance with the mean of predicted values for all instances. The *Shapley Values* for that feature or player is the average marginal contribution [10].

*SHAP* [7] is the software tool we used to obtain the estimated *Shapley Values* with the name 'SHAP value' in the plots from the package. Due to its ability to explain the outputs, especially the visualization of relative contributions of each feature, we use it to examine and interpret each model we build. For example, for *CatBoost*, Figure 3 shows the feature contributions to a single predication in the 'Drills' category based on *Shapley Values* of all features. We can see that the feature 'Product Weight (lb.)' contributes the most to the change of price for this instance from the average predicted price for all instances and 'Chuck Size' contributes the second most to the change of price. The base value is the average value of all predictions from all instances, and the output value is the prediction for a specific instance. From Figure 3, both 'Product Weight (lb.)' and 'Chuck Size' push the price lower, while the feature 'Product Width (in.) pushes the price higher to approach the base value. The feature contributions for the overall training data in the 'Drills' category is shown in Figure 4 where 'Feature value' in y-axis stands for the overall *Shapley Values* for a specific feature. The mean of *Shapley Values* for each feature is shown in Figure 5. Indicated by the *Shapley Values*, 'Motor Type' is the second most significant feature which means it is strongly correlated to the target variable 'price'. From the two graphs, we can clearly understand how each feature contributes to the price in a quantitative way.

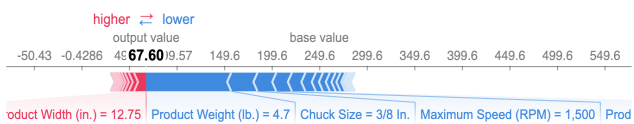

**Figure 3: Feature Contributions for a Single Record**

Finally, we compute the average *Shapley Values* using all *Shapley Values* from above models for each feature and rank them accordingly to get the final feature ranking result for each category. We use the feature directions learned from the above linear model as the final feature signs. The solid theoretical foundation for *Shapley Values* to explain the feature contributions in prediction and our exploration based on rich visualization of *SHAP* guide us to design such a final ranking method.

## 4 EXPERIMENTS AND RESULTS

In this section, we describe a strong baseline algorithm named *Left Navigation Algorithm* that our proposed approach was compared with. In Section 3, we describe the process to calculate the final feature ranking per category using machine learning models with *Shapley Values*. We name our approach *Machine Learning Approach*. We discuss the experimental setup and the metrics for the offline evaluation as well as for the online A/B tests. The detailed results of the offline evaluation and the online A/B tests are also shown and discussed in this section. For offline evaluation, we do not find exact similar open datasets with detailed product attributes and prices so we just use our production data for evaluation.

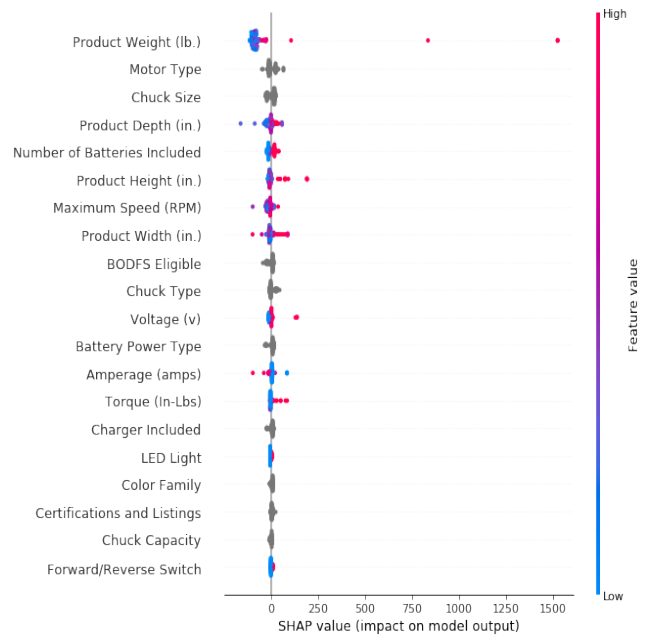

**Figure 4: Visualization of all Shapley Values of Training Data for Feature Ranking**

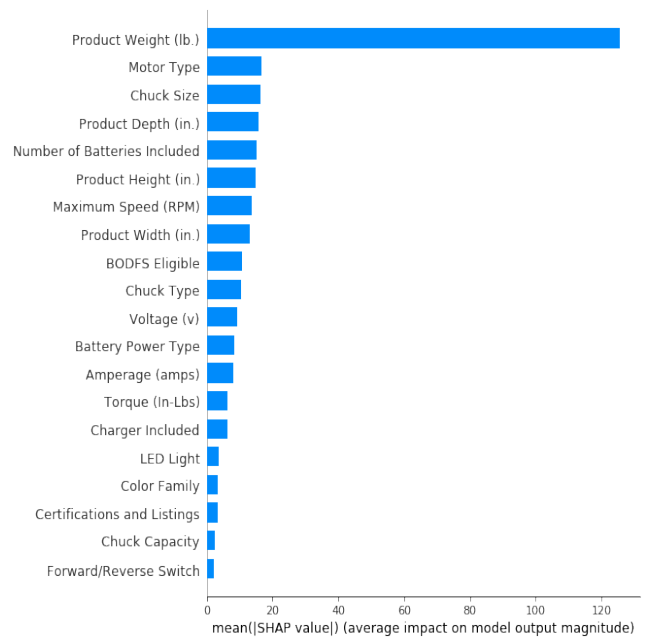

**Figure 5: Feature Ranking based on Mean of Shapley values for Training Data**

## 4.1 Baseline Algorithm

The baseline algorithm is called *Left Navigation Algorithm* powered by customer click stream data. The left navigation features are listed on the left side of a product list page for a leaf node category to let customer select the suitable feature values. An example of the 'Refrigerator' category is shown in Figure 6.

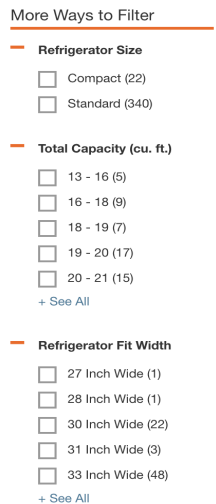

**Figure 6: Left Navigation Features from Customer Behavior**

The *Left Navigation Algorithm* is based on the customer engagement with left navigation feature frequencies. Certain categories tend to have rich left navigation interaction data and certain categories have none/little interaction data. The algorithm uses the engaged feature frequency from the left navigation to rank the feature importance. This method suffers from 1) the cold start problem: if there is little or no features on the left navigation, there is no way to obtain the engaged frequency; and 2) the error perpetuation problem which means customers will only get left navigation interactions for features that we show on the page and cannot get the engagement for the relevant features not shown in the left navigation for customers to begin with. Even this algorithm has drawbacks, it is still a strong baseline since it partially reflects the customer preference.

## 4.2 Experiment Setup

There are many interesting categories in our product catalog. We randomly choose 7 categories to conduct the experiments including Drills, Bathroom Vanity with Tops, French Door Refrigerators, Ceiling Fans, Dehumidifiers, Built-in Dishwashers, Portable Generators. For each category, the baseline algorithm provides the key product features. Our proposed machine learning approach runs through the ML pipeline and generates the ranking features for each category. We select a range of most important hyper-parameters for *LightGBM* and *CatBoost*. For Linear Regression, we use the default settings. For *LightGBM*, the main hyper-parameters include learning rate with 0.04, 0.05, 0.06, 0.09 and number of leaves with 25, 30, 35, 40. For *CatBoost*, the main hyper-parameters include learning rate with 0.10, 0.15, 0.20, 0.25 and depth with 3, 6, 9, 12. With grid search along with cross validation, ML algorithms in our pipeline produce their best validation scores with the optimal hyper-parameters. We use RMSE as the cross validation loss. Each category has its own optimal hyper-parameters from the settings.

## 4.3 Offline Evaluation

For the offline evaluation, we leverage our internal experts with domain knowledge to extract key product features for the specific categories we selected in the test. Based on their expertise, labels were created for each category which are a ranked list of important features. We evaluate *Left Navigation Algorithm* and our proposed *Machine Learning Approach* based on metrics such as NDCG, precision and recall for top 5 and top 10 features. We also evaluate both algorithms for the percentage of the categories that the algorithm can cover because we are interested in expanding the feature recommendations to all the possible categories and products. The performance is shown in Table 1.

- **NDCG, Precision and Recall**: From Table 1, our approach has better NDCG scores for 5 categories including 'Bathroom Vanities with Tops', 'Ceiling Fans', 'Dehumidifiers', 'Built-In Dishwashers' and 'Portable Generators'. For 'Drills', our approach has a close NDCG with the baseline. The baseline has a better NDCG for 'French Door Refrigerators' category. Our proposed approach has higher recall and precision in general for both the top 5 and top 10 features across the sampled categories. This is because the initial feature selection for the *Left Navigation Algorithm* is biased as it does not consider the price. On the other hand, our approach uses the product price as the label to train the models and let the machine learning models learn the most important features that drive the price difference. To be specific, for precision@5, the baseline works well and even better than our proposed approach. For example, 'Built-In Dishwashers' has better precision@5 score with the baseline than our approach. 'Built-In Dishwashers' and 'Portable Generators' both have better precision@10 scores with the baseline than our approach. The primary reason is because the important features are well understood for those categories with the baseline. However, most of the categories are not explored and well understood so our *Machine Learning Approach* has strong advantages. Even for other well known categories like 'Drills', our approach also shows competitive results. Generally speaking, for popular category such as 'French Door Refrigerators' and 'Drills', the baseline works well. If the category is not popular like 'Ceiling Fans' where there is no efficient way for baseline to know the important features, our ML approach can learn them from product price.

Table 1: Performance of the *Left Nav Algorithm* Versus the *Machine Learning Approach* for Sampled Categories.

| Category | Left Navigation Algorithm | | | | |
|---|---|---|---|---|---|
| | NDCG | Precision@5 | Precision@10 | Recall@5 | Recall@10 |
| Bathroom Vanities with Tops | 0.65 | 0.40 | 0.30 | 0.40 | 0.60 |
| Drills | 0.76 | 0.40 | 0.30 | 0.40 | 0.43 |
| French Door Refrigerators | 0.62 | 0.25 | 0.10 | 0.25 | 0.25 |
| Ceiling Fans | 0.43 | 0.20 | 0.10 | 0.10 | 0.20 |
| Dehumidifiers | 0.67 | 0.40 | 0.30 | 0.40 | 0.43 |
| Built-In Dishwashers | 0.53 | 0.35 | 0.20 | 0.20 | 0.40 |
| Portable Generators | 0.81 | 0.40 | 0.25 | 0.20 | 0.40 |
| Category | Machine Learning Approach | | | | |
| | NDCG | Precision@5 | Precision@10 | Recall@5 | Recall@10 |
| Bathroom Vanities with Tops | 0.72 | 0.60 | 0.30 | 0.60 | 0.76 |
| Drills | 0.74 | 0.40 | 0.40 | 0.60 | 0.71 |
| French Door Refrigerators | 0.54 | 0.50 | 0.40 | 0.60 | 0.71 |
| Ceiling Fans | 0.89 | 0.40 | 0.30 | 0.50 | 0.75 |
| Dehumidifiers | 0.78 | 0.80 | 0.60 | 0.57 | 0.86 |
| Built-In Dishwashers | 0.63 | 0.20 | 0.15 | 0.37 | 0.46 |
| Portable Generators | 0.91 | 0.40 | 0.15 | 0.40 | 0.75 |

- **Coverage**: From our product catalog, we would like to cover more categories and products with our feature recommendations. Compared with the current *Left Navigation Algorithm*, our proposed *Machine Learning Approach* increases the category coverage by **45%**. For the categories that do not have feature recommendations because of the cold start problem in the *Left Navigation Algorithm* without customer engagement, our *Machine Learning Approach* can provide the feature recommendations from scratch using product features and prices. This is very successful for our business to significantly expand the feature recommendations.

Due to the time and space limitation, we are not providing the validation scores for each raw feature rankings from the linear and tree models compared with the baseline, and the validation scores of the Shapley version of the feature ranking for each model. We have several insights to share here: 1) the feature ranking using *Shapley Values* generally works better than the original feature ranking for each model we have built; 2) for categorical features, their rankings from the models that directly handle those features are better than the rankings from the models using one-hot encoding to split them and later make the summation of the importance scores from the binary features.

## 4.4 Online A/B Tests

Two online A/B tests are launched to finally evaluate the effectiveness of our proposed machine learning approach compared with the baseline. We run the test on the sampled categories. Each test ran for three weeks to gather sufficient data from user traffic to determine whether the effect seen was statistically significant (or not). We use *Conversion Rate* as the primary metric for the evaluation. *Conversion Rate* is defined here as the number of purchases divided by number of visits. This is used to measure the relevancy of the related algorithms. The rational to use *Conversion Rate* is because it captures the customer purchase behaviors given the control and test experiences. Higher *Conversion Rate* means that customers find the recommended features more useful to help them figure out the right products they need, and vice versa.

*4.4.1 A/B Test for Product Feature Recommendation.* This is the first use case on our e-commerce site which shows the most important features to differentiate the product candidates and better fit the customers' needs. Our *Machine Learning Approach* has a comparable performance against the *Left Navigation Algorithm* which will have a strong positive impact on our business with a broader coverage given the large volume of sales.

*4.4.2 A/B Test for Explainable Recommendation using Feature Message.* This is the second use case on our e-commerce site which shows a message detailing the single most important feature impacting the price difference from amongst the alternative products when compared with an anchor product which is defined as the main product on that product information page. For example, customers see a message right below the product price stating 'Upgrade to 4 burners for $130 more' or 'Spend $300 more for 6 burners' when buying a grill. They then have a better understanding about how they can spend more money to get more sophisticated features or spend less money to get less sophisticated features. From the A/B test result, our *Machine Learning Approach* also has a comparable performance against the baseline algorithm which provides a coverage increase for our business.

## 4.5 Production Impact

Both offline evaluation and online A/B tests show promising results when using our *Machine Learning Approach* in place of the *Left Navigation Algorithm* baseline. We scale up from the sample categories to most of leaf node categories on our production site. Each leaf node category will have its own models trained with best hyper-parameters and the optimal feature ranking list to serve our customers. The category coverage is significantly increased using the proposed approach, we expect this approach to have a noticeable impact on our conversion rates and drive noticeable business value.

## 5 CONCLUSION

Online product feature recommendations help customers purchase the best product between alternative products based on important features with the right values. We proposed interpretable machine learning methods to determine the key product features for customers to help them differentiate the most suitable products. We identify this unique product feature recommendation problem from a business perspective. We formulate the problem as a price-driven supervised machine learning problem to discover the product features that best explain the price of a product in a given product category. We leverage linear model and tree models with *Shapley Values* to rank and recommend the most essential features. We also leverage human experts to evaluate the relevancy of these recommendations when compared to a strong baseline that is based on customer behavior. The offline evaluation shows that our method is superior to the baseline. Our proposed method also shows a comparable conversion rate against the baseline in online A/B tests.